\documentclass[aps,prb,twocolumn,showpacs,superscriptaddress]{revtex4}
\usepackage{amssymb,amsmath}
\usepackage{graphicx}
\usepackage{dcolumn}
\usepackage{subfigure}
\usepackage{bm}

\begin{document}

\title{Elastic Constants of Incommensurate Solid $^{\bf 4}$He}

\author{Claudio Cazorla}
\affiliation{Institut de Ci$\grave{e}$ncia de Materials de 
             Barcelona (ICMAB-CSIC), 08193 Bellaterra, Spain}
\author{Yaroslav  Lutsyshyn}
\affiliation{Institut f\"{u}r Physik, Universit\"{a}t Rostock, 
             18051 Rostock, Germany}
\author{Jordi Boronat}
\affiliation{Departament de F\'{i}sica i Enginyeria Nuclear, 
             Universitat Polit\`{e}cnica de Catalunya, Campus 
             Nord B4-B5, E-08034, Barcelona, Spain}
\email{ccazorla@icmab.es}

\begin{abstract}
We study the elastic properties of incommensurate solid $^{4}$He in the limit of zero temperature.
Specifically, we calculate the pressure dependence of the five elastic constants 
($C_{11}$, $C_{12}$, $C_{13}$, $C_{33}$, and $C_{44}$), longitudinal and transversal speeds of 
sound, and the $T = 0$ Debye temperature of incommensurate and commensurate hcp $^4$He using the 
diffusion Monte Carlo method. 
Our results show that under compression the commensurate crystal is globally 
stiffer than the incommensurate, however at pressures close to melting (i.e. $P \sim 25$~bars) 
some of the elastic constants accounting for strain deformations of the hcp basal 
plane ($C_{12}$ and $C_{13}$) are slightly larger in the incommensurate solid. 
Also, we find that upon the introduction of tiny concentrations of point defects  
the shear modulus of $^{4}$He ($C_{44}$) undergoes a small reduction. 
\end{abstract}
\maketitle

\section{Introduction}
\label{sec:introduction}

An intriguing resemblance between the dependence of the shear modulus (SM) 
and torsional oscillator (TO) frequency changes on temperature, amplitude, 
and concentration of $^{3}$He impurities, has been experimentally observed 
in solid $^{4}$He at low temperatures.~\cite{day07}  
Crystal defects are clearly involved in both phenomena, however how SM and 
TO fluctuations are exactly related remains yet a puzzle. Day and Beamish 
identified the stiffening of solid helium with decreasing temperature, 
i.e., increase of its shear modulus, with the pinning/unpinning of dislocations 
by isotopic impurities. Subsequent experiments have confirmed 
Day and Beamish interpretations~\cite{day09} although recent elasticity 
measurements on ultrapure single crystals seem to suggest that SM variations 
cannot be uniquely understood in terms of mobile dislocations.~\cite{rojas09,rojas10}  

Torsional oscillator anomalies were first interpreted  
as the mass decoupling of a certain supersolid fraction,~\cite{kim04a,kim04b} 
a counterintuitive physical phenomenon that Andreev and Lifshitz already proposed  
in solid helium more than 40 years ago.~\cite{andreev69} Supporting 
this view is the fact that TO anomalies appear to occur only in bulk 
$^{4}$He.~\cite{kim08} Nevertheless, the supersolid interpretation of TO 
anomalies appears to leave open its connection to SM fluctuations and 
diverse theoretical arguments and hypotheses have been put forward in an attempt 
to simultaneously rationalize the origins of both anomalies.  
Anderson, for example, proposes that supersolidity is an intrinsic property of 
bosonic crystals, which is only enhanced by disorder, and that the elastic anomaly 
is due to the generation of vortices at temperatures close to the supersolid 
transition.~\cite{anderson} From a diametrically opposite standpoint, Reppy 
has argued that the TO behavior is caused by an increase of the $^{4}$He shear 
modulus which mimics mass decoupling by stiffening the TO setup.~\cite{reppy10} 
Other scenarios somewhat more reconciling with the original TO and SM interpretations 
have been also proposed in which for instance mass superflow is assumed to occur in 
the core of dislocations only when these are static.~\cite{balibar10,rojas10}  

As it can be appreciated, definitive conclusions on the roots of SM and TO anomalies 
remain contentious. In a recent paper, Chan \emph{et al.}~\cite{chan12} have shown 
that for solid $^4$He in vycor the nonclassical moment of inertia (NCRI) disappears 
if the TO setup is designed in such a way that is completely free from any shear modulus 
stiffening effect. This result seems to show that NCRI can be
totally attributed to elastic effects and not to the existence of a 
supersolid fraction.~\cite{balibar2}  
On the other hand, a recent experiment in which DC rotation was superposed to both TO and 
SM measures suggested that the cause of both anomalies below a critical temperature 
could have different microscopic origins.~\cite{kimdc} Also, the source of a small 
peak in the specific heat of $^{4}$He~\cite{lin} at temperatures close to that at 
which TO and SM anomalies appear remains yet unexplained.

\begin{figure}[t]
\centerline{
\includegraphics[width=1.00\linewidth]{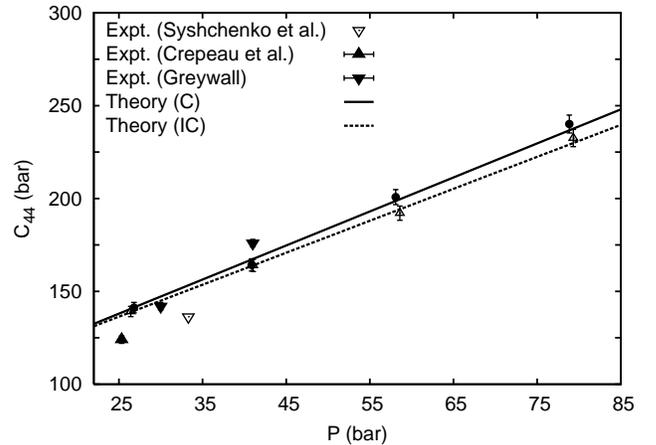}}
\caption{Shear modulus results obtained for C and IC  
         hcp $^{4}$He expressed as a function of pressure. 
         Experimental data from Refs.~\onlinecite{syshchenko09}, 
         \onlinecite{crepeau71}, and \onlinecite{greywall77} are shown for comparison.
         Solid lines represent linear fits to DMC results (see text).}
\label{fig1}
\end{figure}

In this work, we study the change in the elastic constants of solid $^4$He caused by 
the presence of small point defects concentrations, $n_{v}$, of $0.5-2.0$~\%~. 
As it has been shown, the presence of vacancies  
induces a finite superfluid fraction in the crystal (incommensurate crystal, IC) so 
that we can theoretically compare the elastic constants of a supersolid with those of the 
perfect crystal (commensurate crystal, C).  
In particular, we estimate the pressure dependence of the elastic constants 
$C_{ij}$'s ($C_{11}$, $C_{12}$, $C_{13}$, $C_{33}$ and $C_{44}$, where the last one is 
also known as the shear modulus) and derived quantities (the $T=0$ Debye temperature
and transverse/longitudinal speeds of sound) of bulk IC and C hcp $^{4}$He. 
Our calculations show that (i)~under moderate and large compressions the   
C phase is globally stiffer than the IC solid, (ii)~at pressures close to 
melting (i. e. $P \sim 25$~bars) some of the elastic constants accounting for specific 
strain deformations of the hcp basal-plane ($C_{12}$ and $C_{13}$) are slightly larger 
in the IC crystal, and (iii)~the shear modulus difference between C and IC $^{4}$He crystals is 
about $10$ to $90$ times smaller (in absolute value) than the experimentally observed 
$C_{44}$ variation caused by the pinning/unpinning of dislocations.

The remainder of this article is organized as follows. In the next section, we briefly describe the
computational methods employed and provide the details of our calculations. Next, we 
present and discuss the results obtained and summarize the main conclusions 
in Sec.~\ref{sec:conclusions}.   

\begin{figure}
\centerline{
\includegraphics[width=1.00\linewidth]{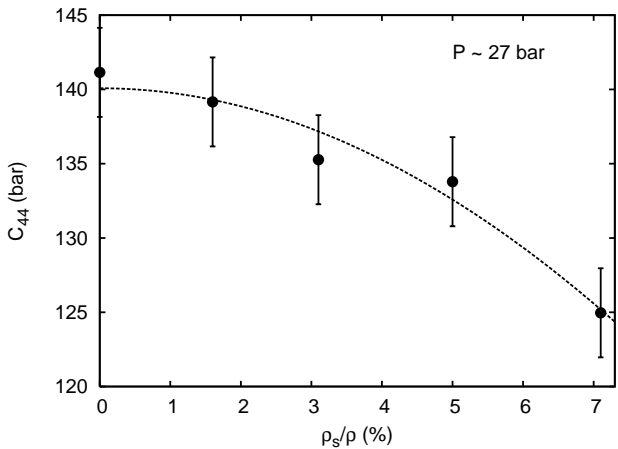}}
\caption{The shear modulus of IC solid hcp $^{4}$He expressed as a function 
        of the superfluid fraction at pressures close to melting. The dashed line is 
        a guide to the eyes.}
\label{fig2}
\end{figure}

\section{Computational Method}
\label{sec:methods}

In this study we employ the diffusion Monte Carlo method (DMC), an accurate
ground-state approach in which the Schr\"odinger equation of a 
$N$-particle interacting system is solved stochastically by simulating
branching and diffusion processes in imaginary time.~\cite{boronat94} 
As it is usual in DMC, we introduce a guiding wave function (gwf) for importance
sampling that crucially reduces the variance of the statistical estimations.     
Our gwf model is symmetric under the exchange of atoms 
and correctly reproduces the experimental equation of state of solid $^4$He and other 
quantum crystals.~\cite{cazorla09,cazorla08,cazorla10} We note that DMC energies 
are virtually exact, i.e. are only subjected to statistical bias, and ultimately 
do not depend on the particular choice of the guiding wave function. 
The value of all technical parameters, i.e. size of the simulation box, 
population of walkers, and length of the imaginary time-step, 
have been set in order to ensure convergence of the total ground-state 
energy to less than $0.01$~K/atom. As in previous works, we modeled the 
$^{4}$He-$^{4}$He interactions with the Aziz~II pairwise potential.~\cite{aziz}    
Further technical details of our elastic constant calculations can be found in 
Refs.~\onlinecite{cazorla12a,cazorla12b}. 

It is important to stress that DMC $C_{ij}$ estimations essentially rely on computation
of total energies as a function of strain thus numerical errors stemming from finite-size 
effects can already be made negligible (i.e. smaller than $0.01$~K/atom) in computationally 
affordable simulation boxes of $24.2$~\AA~$\times$~$24.2$~\AA~$\times$~$27.4$~\AA~ 
containing $200$ atoms.
The IC phase is built by introducing small vacancy concentrations of $0.5-2.0$~\%
in the crystal. Although it is well-known that the presence of point defects in solid $^{4}$He 
is energetically penalized,~\cite{ceperley04} this route allows for 
simulation of supersolids under tight and controllable 
conditions.~\cite{cazorla12b,rota12,yaros10}
 
\begin{figure}
\centerline{
\includegraphics[width=1.00\linewidth]{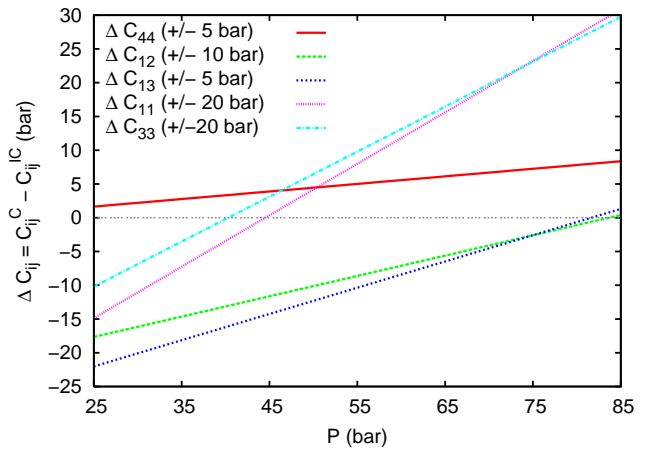}}
\caption{Elastic constant differences between C and IC crystals
        ($\rho_{s}/\rho = 2$~\%) of hcp $^{4}$He expressed as a function of pressure.  
         The size of the error bars are indicated within the parentheses.
         Positive $\Delta C_{ij}$ values indicate softening of the corresponding 
         elastic constant in an hypothetical vacancy-induced normal-to-supersolid phase transition.}
\label{fig3}
\end{figure}

\section{Results and Discussion}
\label{sec:results}

In Fig.~\ref{fig1}, we show the shear modulus of C and IC 
(with $\rho_{s}/\rho = 2$~\%) hcp $^{4}$He expressed as a function of pressure. 
We find that in both states $C_{44}$ behaves linearly with pressure over all the 
range of densities considered, i.e. $0.028 \le \rho \le 0.033$~\AA$^{-3}$~. As 
one may also see, the shear modulus of the C crystal is larger than 
that of the IC solid and the value of the $\Delta C_{44} \equiv C_{44}^{\rm C} - C_{44}^{\rm IC}$ 
difference increases under compression. It must be noted that 
the numerical uncertainty in our $C_{44}$ calculations is $5$~bar thus the 
predicted $\Delta C_{44}$ values are rigorously different from zero at 
pressures above $50$~bar (see Fig~\ref{fig3}). 

In principle, one may expect that besides pressure 
$\Delta C_{44}$ variations are also dependent on the imposed fraction of 
mass superflow, or conversely, the concentration of point defects. 
However, as we show in Fig~\ref{fig2}, such a dependence turns out to 
be rather weak. For instance, in the $0 \le \rho_{s}/\rho \le 3$~\% interval $C_{44}$  
decreases in less than the $5$~\% of its ground-state value and 
even when an excessive $\rho_{s}/\rho$ value of $7$~\% is constrained the 
accompanying variation of shear modulus is of just $\sim -11$~\%.  
Concerning possible temperature effects, it is well-known that the contribution 
of phonon excitations to the thermal energy of solids reduces the speeds of sound by an 
amount that is proportional to $T^{4}$, so implying a $\propto T^{8}$ dependence in the
elastic constants.~\cite{beamish12} Our zero-temperature 
conclusions on $\Delta C_{44}$, therefore, can be fairly generalized to the regime of 
ultralow temperatures (that is, few mK). 
In fact, the ground-state results reported in this study are in very good 
agreement with those obtained by Pessoa \emph{et al.} for hcp $^{4}$He at $T = 1$~K 
using the path integral Monte Carlo method.~\cite{pessoa10,ardila11}     

It must be stressed that our $C_{44}$ results are obtained for pure, i.e. with zero
concentration of $^{3}$He atoms, and free-of-dislocations $^{4}$He single crystals hence 
direct comparisons to Day and Beamish~\cite{day07,day09,syshchenko09} data obtained 
in polycrystals turn out to be very complicate. In the light of our results,
however, one may notice that experimentally observed shear modulus variations caused by 
the pinning/unpinning of dislocations are of opposite sign and about one order 
of magnitude larger ($10-20$~\% in polycrystals and 
$\sim 50-90$~\% in monocrystals~\cite{fefferman12}) than fluctuations reported 
here for hypothetical superfluid mass flows of $\sim 1$~\% (see Fig~\ref{fig2}). 
Consequently, we may conclude that if a vacancy-induced normal-to-supersolid phase 
transition occurred in solid helium dislocation-mediated mechanical contributions to $C_{44}$ 
would totally overwhelm those stemming from mass superflow. Interestingly, 
Rojas \emph{et al.} have recently reported an anomalous 
softening of high quality ultrapure monocrystals in the temperature 
region wherein supersolidity could occur.~\cite{rojas10} 

We have also determined the $\Delta C_{ij}(P)~\lbrace ij = 11, 12, 13, 33 \rbrace$
deviations describing the response of C and IC hcp crystals to strain 
basal plane deformations.~\cite{cazorla12a} 
First, we note that all these components also present a linear dependence on 
pressure (see Fig.~\ref{fig3}, where numerical uncertainties are indicated within 
parentheses). 
Second, $\Delta C_{ij}$ slopes are all positive thus implying that beyond a 
certain critical pressure C hcp $^{4}$He is plainly stiffer than the 
IC crystal. According to our calculations, this critical pressure is above $85$~bar. 
Interestingly, $C_{12}$ and $C_{13}$ are largest, by a small amount, in the 
IC solid at pressures below $50$ and $70$~bar respectively. 
This outcome comes to show that in an hypothetical low pressure 
normal-to-supersolid phase transition, the final supersolid could behave more rigidly 
than the initial normal state under certain strain deformations. 
Nevertheless we find that the $C_{66}$ coefficient, which is defined as 
$\frac{1}{2}\cdot \left(C_{11} - C_{12}\right)$ and can be directly measured 
in acoustic experiments, is always smaller in the IC state.
This behavior is analogous to the tendency found for the shear modulus
although $C_{66}$ variations are in general larger (e.g. at $P = 25$~bar 
$\Delta C_{66} \approx \Delta C_{44}$ whereas at $P = 85$~bar 
$\Delta C_{66} \approx 2 \cdot \Delta C_{44} $).  

\begin{figure} 
\centerline
        {\includegraphics[width=1.0\linewidth]{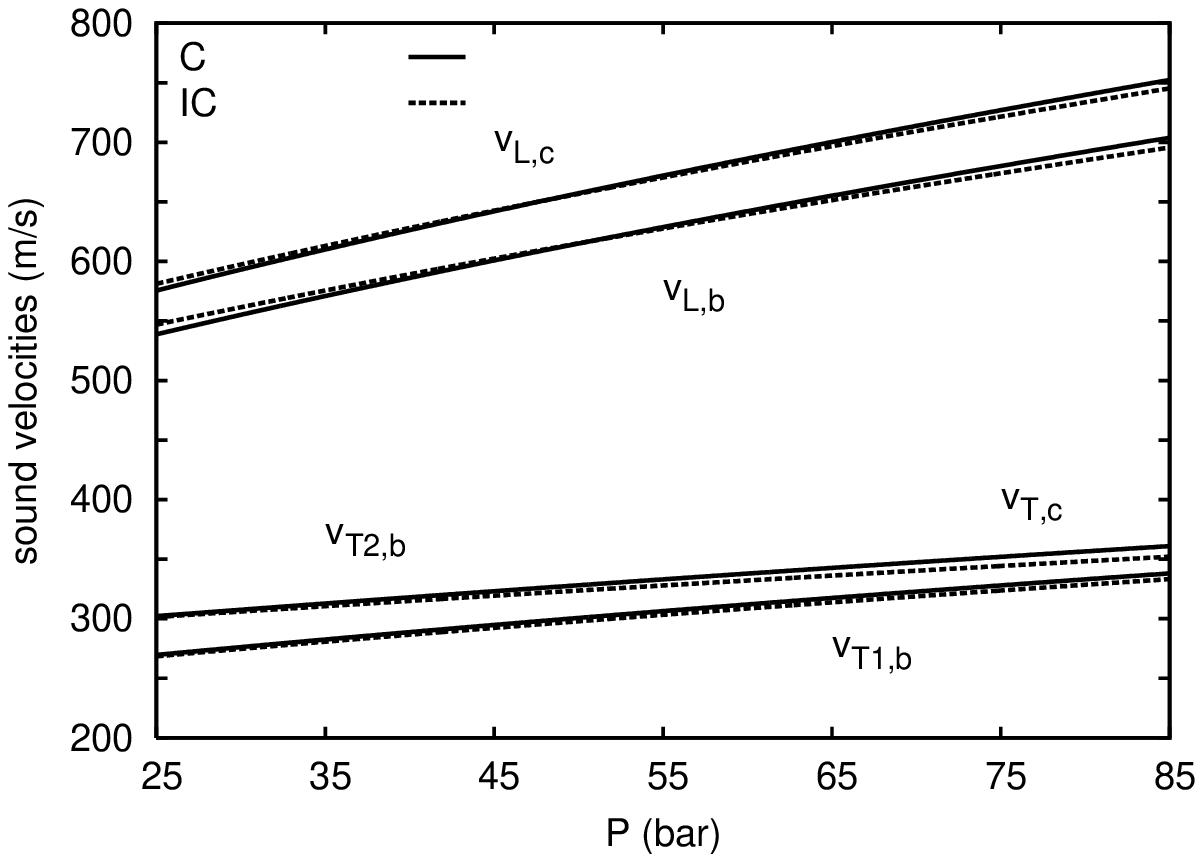}}%
        {\includegraphics[width=1.0\linewidth]{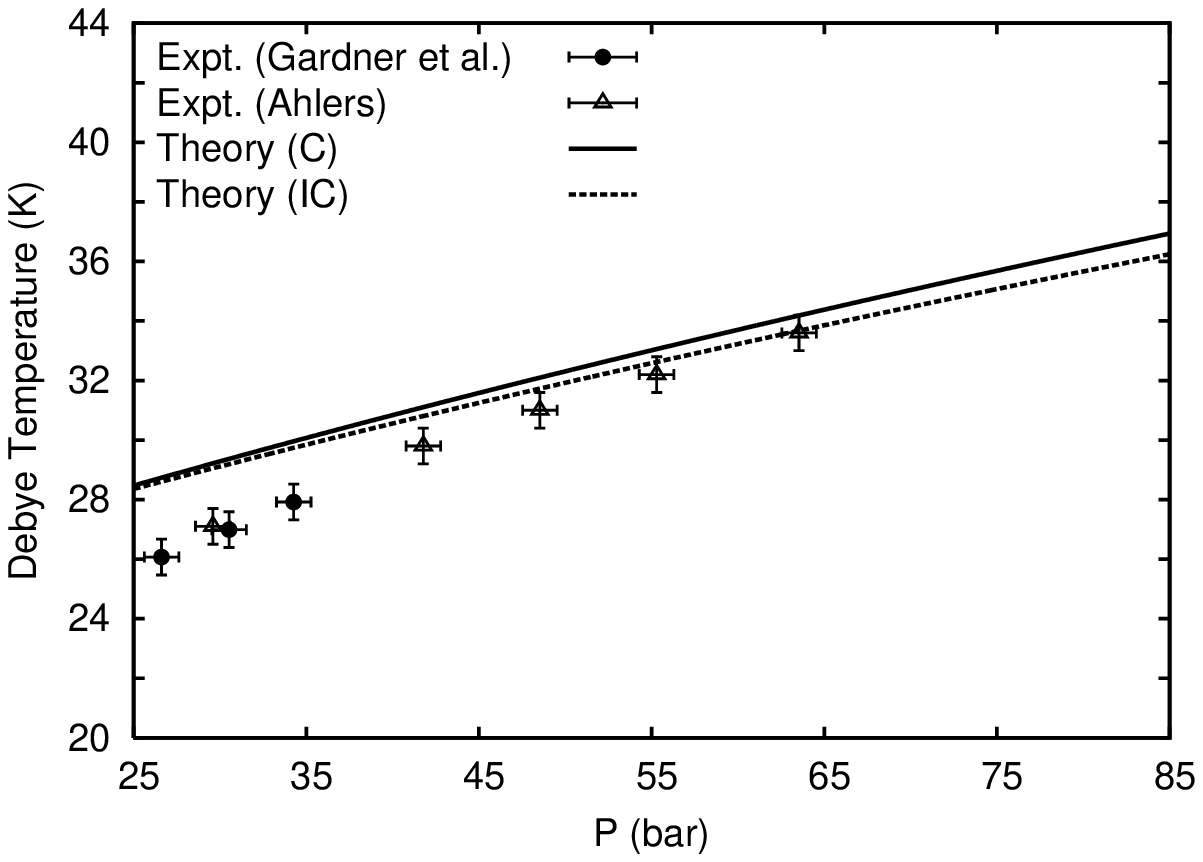}}%
        \caption{\emph{Top:} Calculated longitudinal~(L) and transverse~(T)  
                 speeds of sound  along the basal plane~(b) and $c$-axis~(c)
                 of C and IC ($\rho_{s}/\rho = 2$~\%) hcp $^{4}$He as 
                 a function of pressure.             
                 \emph{Bottom:} Estimated $T = 0$ Debye Temperature of C and IC
                 hcp $^{4}$He as a function of pressure. Experimental data from
                 Refs.~\onlinecite{gardner73} and~\onlinecite{ahlers70} are shown for 
                 comparison.} 
\label{fig4}
\end{figure}

Finally, Fig.~\ref{fig4} shows the calculated longitudinal and transversal speeds of sound  
($v_{L}$ and $v_{T}$) of C and IC hcp $^{4}$He under pressure.~\cite{cazorla12a}
As one can observe, $v_{L}$ velocities along the hcp $c$-axis and basal plane 
are slightly larger in the IC crystal within approximately 
the same pressure range in which $\Delta C_{12}$ and $\Delta C_{13}$ deviations 
are found to be negative. Nevertheless, speeds of sound deviations near melting 
turn out to be so small that in practice these could probably not be detected with standard means. 
The same can be concluded about the $T = 0$ Debye temperature for which, as we show 
in Fig.~\ref{fig4}, the corresponding C-to-IC variation is smaller than 
the typical experimental precision. In view of these technical limitations, it would be 
very interesting to perform new $C_{ij}$ and $v_{L,T}$ measurements on $^{4}$He at large 
pressures (i.e. $P \ge 60$~bar) where C-to-IC differences develop larger. 
To this regard, spectroscopic measurements of the $E_{2g}$ phonon mode (i.e. the shear 
mode corresponding to the beating of the two hcp sublattices against each other in the
two orthogonal directions of the basal plane) would be particularly helpful   
since in this type of experiments (i)~$C_{ij}$ values can be 
determined with a very small imprecision of less than the $2$~\%, 
(ii)~tiny solid samples are needed (i.e. of $\mu$m size) thus likely crystal 
quality issues present in SM and TO experiments could be somehow alleviated, and 
(iii)~pressure conditions can be efficiently tuned.~\cite{eckert78}

\section{Conclusions}
\label{sec:conclusions}

To summarize, we have studied the elastic properties of hcp solid $^{4}$He in 
a metastable IC state and compared them to those obtained for its 
C ground-state.
Our calculations show that near melting elastic constants $C_{11}$ and $C_{12}$ 
accounting for specific strain deformations of the hcp basal plane are slightly larger 
in the IC crystal. At moderate and high pressures,
however, the C phase is always stiffer than the IC. 
Also, we find that the appearance of a finite superfluid fraction  
(e.g. $\rho_{s}/\rho \sim 1$~\%) caused by the introduction of vacancies  
unequivocally provokes a small decrease of the $^{4}$He shear modulus (i.e. $\Delta C_{44} \sim 1$~\%). 
We argue then that if a vacancy-induced normal-to-supersolid phase transition occurred in helium 
crystals containing isotopic impurities and line defects, dislocation-mediated contributions 
to $C_{44}$ would totally overwhelm those stemming from mass superflow. 
As an alternative to usual dynamic experiments focused on the search of hypothetical supersolid
manifestations, we suggest to perform spectroscopic measurements of the $E_{2g}$ mode 
of $^{4}$He at moderate and high pressures. 

\begin{acknowledgments}

This work was supported by MICINN-Spain Grants No. MAT2010-18113,
CSD2007-00041, and FIS2011-25275, and Generalitat de Catalunya 
Grant No.~2009SGR-1003,

\end{acknowledgments}

\end{document}